\documentclass[12pt,a4paper,english,superscriptaddress, aps, preprint]{revtex4}
\usepackage{amsmath}
\usepackage{amssymb}
\usepackage{graphicx}
\usepackage{hhline}
\usepackage{mwe}
\usepackage{amsbsy}
\usepackage{textcomp}
\usepackage{commath}
\usepackage{subfig}

\usepackage{stackengine}
\stackMath

\makeatletter
\usepackage{babel}
\newcommand{\bea}{\begin{eqnarray}}
\newcommand{\eea}{\end{eqnarray}}

\newcommand{\pa}{\partial}

\newcommand{\be}{\begin{equation}}
\newcommand{\ee}{\end{equation}}
\numberwithin{equation}{section}

\begin{document}
\immediate\write16{<<WARNING: LINEDRAW macros work with emTeX-dvivers
                    and other drivers supporting emTeX \special's
                    (dviscr, dvihplj, dvidot, dvips, dviwin, etc.) >>}

\title{ Moduli spaces of BPS lumps with holomorphic impurities}
\date{03-12-2021}
\author{J. M. Queiruga}
\email{xose.queiruga@usal.es}
\affiliation{Department of Applied Mathematics, University of Salamanca,
37008, Salamanca, Spain}
\affiliation{Institute of Fundamental Physics and Mathematics, University of Salamanca, 37008 Salamanca, Spain}
\affiliation{Department of Physics, UPV/EHU,
48080, Bilbao, Spain}

\begin{abstract}
A self-dual generalization of the lump-impurity system is introduced. This model possesses lump-antilump-like pairs as static solutions of the pertinent Bogomolny equations. This allows for a moduli space approximation analysis of the BPS solutions which are identified as lump-antilump configurations. Some geometrical properties of the resulting moduli are analyzed. In addition, it is argued that, this type of impurity models can be interpreted as a limit of certain non-impurity theories. 

  \end{abstract}

\maketitle

\section{Introduction}

The study of soliton-antisoliton annihilation processes in nonlinear field theories entails many difficulties. One of the main reasons for this is that, in general, a soliton-antisoliton configuration does not belong to any BPS sector. This implies, in particular, that there is always a non-vanishing static interaction between solitons of opposite charges and as a consequence, the canonical moduli space approximation \cite{moduli-1}, is not available. In general, the study of such processes requires a full numerical analysis, which often does not allow to extract sufficiently general properties.  

It was realized \cite{hook, tong} that it is possible to preserve (part of) the BPS structure in models where a non-dynamical field (impurity) is introduced (for recent advances see \cite{ vortex-imp-2, BPS-imp-phi4, vortex-imp-3, no-force, solvable-imp, queiruga_skyrme_imp, kim}). As in standard BPS theories, there is a moduli space of solutions, such that trajectories over this internal space connect isoenergetic configurations. If one considers only translational degrees of freedom, the shape of the soliton does not change through the moduli space in the standard case. Once the impurity is introduced, the shape of the soliton gets distorted close to the impurity, in such a way that the combination of this deformation with the translation still preserves the BPS property. Moreover, each point in the moduli space is associated to different spectral structures, and this feature leads to very interesting phenomena, like for example the so-called spectral wall \cite{ spec-wall}. In addition, the fact that the full system preserves a part of the BPS structure implies the absence of static soliton-impurity interaction and this allows for a clean study of internal modes in multikink collisions.

It was soon understood in one-dimensional models that for a specific choice, the impurity can generate a sort of kink solutions which can be interpreted as kink-antikink configurations. The moduli space of such configurations is still one-dimensional, as the only degree of freedom is associated to the generalized translational symmetry (the impurity is non-dynamical and does not contribute with degrees of freedom to the moduli). In contrast to standard situations, even the moduli associated to a single soliton in impurity models is nontrivial and effectively describes the soliton-impurity interaction. Beyond one-dimensional models, there is a rather poor understanding of soliton-antisoliton processes, both at analytical and numerical level. In this sense, this work presents a new approach to study some properties of soliton-antisoliton configurations that preserve the BPS structure in a $2+1$ dimensional model.

 We will focus on a modification of the $\mathbb{C}P^1$ model, whose BPS solutions are rational maps between two-dimensional Riemann spheres, either holomorphic or antiholomorphic. The impurity field will be introduced in such a way that it can be interpreted as a frozen lump with opposite topological charge with respect to the dynamical lump.  Of course, if one wants to preserve part of the BPS structure (and therefore the canonical moduli space), the coupling between impurity and dynamical field (the lump) needs to have a particular form, dictated by SUSY. As it is well-known, the underlying supersymmetric structure can be used to obtain the Bogomolny equations necessary to build the moduli space.  We will study some properties of such BPS solutions and the moduli spaces obtained from them.

This paper is organized as follows. In Sec. 2 we construct the BPS impurity Lagrangian and compute BPS equations and BPS bound using SUSY techniques. In Sec. 3, general properties of the solutions as well as some connection to other models found in the literature as discussed. In Sec. 4 we describe certain properties of the moduli space and present some examples. In Sec. 5, geometrical properties of the impurity moduli are analyzed, namely, the Kähler structure. Sec. 6 contains some considerations about the relation between impurity models and some multifield systems. Finally, Sec. 7 is devoted to our conclusions and further discussion. We have also added an Appendix with the explicit calculation of the degree formula of the BPS solutions for general impurities.


\section{Lagrangian and BPS equations}
\label{Lagrangian}

Perhaps, the simplest nonlinear $\sigma$-model in two dimensions is the $O(3)$ model \cite{MS}. It can be formulated in terms of a unit 3-vector and its Lagrangian is given by a Dirichlet term plus a constraint
\be
\mathcal{L}_{O(3)}= \frac{1}{4}\partial_\mu \phi^a \partial^\mu\phi^a, \,\, \phi^a\phi^a=1, \,\, a=1,2,3.
\ee
 In two dimensions, the solutions of the model are maps from the one point compactification of $\mathbb{R}^2$ ($\simeq \mathbb{S}^2$ ) to $\mathbb{S}^2$. There is a geometrically meaningful formulation of the model written in terms of the stereographic projection
\be
\phi^a=\left(\frac{u+\bar{u}}{1+\vert u\vert^2},\frac{-i (u-\bar{u})}{1+\vert u\vert^2},\frac{1-\vert u\vert^2}{1+\vert u\vert^2}\right).
\ee

 This model is usually called the $\mathbb{C} P^1$ model, for obvious reasons. In terms of the complex field $u$ the Lagrangian is given by 
\be
\mathcal{L}_{CP^1}= \frac{1}{2}\frac{ \partial_\mu u \partial^\mu \bar{u}}{(1+|u|^2)^2},
\ee
where we have introduced a $1/2$ factor for later convenience. The addition of a complex background field $\sigma$ to the field $u$, say in the form $\sigma V(u)$, breaks in general the BPS structure enjoyed by the original $\mathbb{C}P^1$ Lagrangian. The idea is to look for specific couplings such that a fraction of the BPS property is preserved. In order to do that, we can use the underlying SUSY structure of the $\sigma$-model. In the off-shell SUSY formulation we have
\be
\mathcal{L}_{\mathbb{C}P^1}= \frac{1}{2}\frac{ \partial_\mu u \partial^\mu \bar{u}+F\bar{F}}{(1+|u|^2)^2}+\text{fermions},
\ee
where the fermionic terms are not displayed explicitly and $F$ is a complex auxiliary field. It is possible to add (in the SUSY formulation of the model) a BPS preserving impurity term of the form
\be
\mathcal{L}_{\text{impurity}}=\frac{1}{(1+\vert u \vert^2)^2}\sigma(z)(F-2\pa_{\bar{z}} u)+\text{h.c.}+\text{fermions}.\label{sub:imp1}.
\ee
It is easy to show that this type of terms preserve half of the fermionic SUSY transformations and, as a consequence, half of the BPS structure \cite{BPS-imp-susy}. Although $\sigma$ may have any dependence on the spatial coordinates without spoiling the BPS structure, we have chosen a holomorphic form. We will see later that, with this choice, the impurity can be interpreted as a sort of frozen lump. By imposing the field equation of the auxiliary field we get 
\be\label{aux}
F=-2\bar{\sigma}, \, \bar{F}=-2\sigma.
\ee
Finally, after substituting the auxiliary field equation into the Lagrangian we have
\be\label{full-lag}
\mathcal{L}=\mathcal{L}_{\mathbb{C}P^1}+\mathcal{L}_{\text{impurity}}=\frac{1}{(1+|u|^2)^2}\left(\frac{1}{2}\partial_\mu u \partial^\mu \bar{u}-2\sigma\bar{\sigma}-2\sigma\pa_{\bar{z}}u-2\bar{\sigma}\pa_z\bar{u}\right).
\ee
 
We can use the Bogomolny rearrangement to decompose the static energy as follows
 \be\label{energ_1}
 E=\int d^2 x \frac{1}{(1+|u|^2)^2}\left(2\vert \pa_{\bar{z}} u +\bar{\sigma}\vert^2+\left(\pa_{z}u\pa_{\bar{z}}\bar{u}-\pa_{\bar{z}} u\pa_{z}\bar{u}\right)\right).
 \ee
 Therefore we have
 \be
 E\geq  \pi \vert Q \vert, \, Q=\frac{1}{\pi}\int d^2x \frac{\pa_z u\pa_{\bar{z}}\bar{u}-\pa_{\bar{z}}u\pa_{z}\bar{u}}{\left(1+ \vert u\vert^2\right)^2}
 \ee

As one can see from (\ref{energ_1}), the (standard) topological bound is saturated when $u$ satisfies the BPS equation
\be
\pa_{\bar{z}}u+\bar{\sigma}=0.\label{eq:BPS}
\ee 

It is a straightforward exercise to verify that (\ref{eq:BPS}) also satisfies the second order Euler-Lagrange equations. This equation can be also obtained directly from the kernel of the SUSY transformation of the fermions in the chiral superfield together with (\ref{aux}). The general form of solutions of (\ref{eq:BPS}) with finite energy is given by
\be
u=\frac{p(z)}{q(z)}-\int \bar{\sigma}d\bar{z}, \label{BPS:sol}
\ee
where $p(z),q(z)$ are holomorphic polynomials. The important feature about (\ref{BPS:sol}) is that it is a combination of a holomorphic function (the lump solution of the original $\mathbb{C}P^1$ model) plus an antiholomorphic function defined by the impurity. This form resembles that of a superposition of a lump and an antilump, which unlike the $\mathbb{C}P^1$ case, belongs to the BPS sector. Even though for concreteness we have focussed our calculations on the $\mathbb{C}P^1$ model, previous results are straightforwardly generalized to any nonlinear $\sigma$-model in two dimensions.


\section{Some properties of the solutions}\label{imp-0-inf}

As it is well-known, the topological degree of a lump is given by the maximum degree of the pair of polynomials that enter in the rational map (assuming that the rational maps are given in the irreducible form). However, the topological degree of the soliton in the presence of the antiholomorphic impurity depends nontrivially on $\sigma$ itself.  We will focus our analysis on impurities of the form
\be\label{imp_1_1}
\int d\bar{z}\sigma(\bar{z})=\frac{\sigma_1(\bar{z})}{\sigma_2(\bar{z})},
\ee
where $\sigma_{1,2}(\bar{z})$ are polynomials. We have chosen to write the impurity as in (\ref{imp_1_1}) in order to write the BPS solution (\ref{BPS:sol}) in a simple way
\be
u=\frac{p(z)}{q(z)}+\frac{\sigma_1(\bar{z})}{\sigma_2(\bar{z})}.\label{sol1}
\ee
 The rational map $R(z)=p(z)/q(z)$ is an arbitrary holomorphic function while the antiholomorphic part is fixed by the impurity.


\subsection{Negative degree impurities} 
 
We now choose the following boundary condition at infinity
\be
\lim_{\vert z\vert \rightarrow \infty} u=0.\label{boundary}
\ee

This condition constrains the relative degree of the polynomials $p, q, \sigma_1, \sigma_2$ as follows
\be
\deg p<\deg q,\,\, \deg \sigma_1 <\deg \sigma_2\,\label{condpol}.
\ee
 i.e., both lump and impurity have a negative degree at infinity. The analysis performed in Appendix A shows that the topological degree of a solution of the form (\ref{sol1}) with the conditions (\ref{condpol}) is 
 \be
 N=\deg q -\deg \sigma_2.
\ee

Therefore, it seems that one can interpret a charge N solution in the presence of this kind of impurities as a configuration of  $\deg q$ lumps determined by the rational map and $\deg \sigma_2$ antilumps determined by the impurity. In order to confirm this interpretation one can compute the sign of the topological charge density at the zeros of $q$ and $\sigma_2$. Let us assume for now, that the polynomials $q$ and $\sigma_2$  have respectively $n_1$ and $n_2$ different roots and no common roots with $p$ and $\sigma_1$,
\bea
q&=&(z-a_1)(z-a_2)...(z-a_{n_1}),\\
\sigma_2&=&(\bar{z}-b_1)(\bar{z}-b_2)...(\bar{z}-b_{n_2}).
\eea

The topological charge density is given by

\be
n=\frac{1}{\pi}\frac{\vert\pa_z u\vert^2-\vert \pa_{\bar{z}}u\vert^2}{(1+\vert u \vert^2)^2},
\ee
which in our case takes the form
\be
n=\frac{1}{\pi}\frac{\vert \Delta_{pq}\vert^2\vert \sigma_2\vert^4-\vert \Delta_{\sigma_1 \sigma_2}\vert^2 \vert q\vert^4}{\left(\vert q\vert^2\vert \sigma_2 \vert^2+\vert \Sigma \vert^2\right)^2},
\ee
where
\bea
\Delta_{pq}&=& q \pa_z p-p \pa_z q,  \\
\Delta_{\sigma_1 \sigma_2}&=& \sigma_2 \pa_{\bar{z}} \sigma_1- \sigma_1 \pa_{\bar{z}} \sigma_2, \\
\Sigma&=&\sigma_2 p+\sigma_1 q.
\eea

Now, close to one of the roots of $q$, say $a_i$ we have
\bea
n=\frac{1}{\pi}\Pi_{j\neq i}\vert a_i-a_j\vert^2\Pi_{j=1}^{n_2}\vert \bar{a}_i-b_j\vert^2>0 ,
\eea
while close to a root of $\sigma_2$, $b_i$ we have
\bea
n=-\frac{1}{\pi}\Pi_{j\neq i}\vert b_i-b_j\vert^2\Pi_{j=1}^{n_1}\vert \bar{b}_i-a_j\vert^2<0 .
\eea
This confirms the interpretation of the solutions (\ref{sol1}) as a configuration of $n_1$ lumps and $n_2$ antilumps. The argument can be extended for other base point conditions, but still the topological charge density will be concentrated about the poles of the rational functions (the zeroes of $q(z)$ and $\sigma_2(\bar{z})$). Let us go back for a moment to the ordinary $\mathbb{C}P^1$ model. Let us consider a  lump described by the following polynomial
\be
u=(z-a_1)(z-a_2)...(z-a_n).
\ee
 If the zeroes of $u$, $a_1, a_2, ...., a_n$ are well separated, this solution can be interpreted as a system of  $n$ lumps, located at the position of the zeroes. In addition, the $\mathbb{C}P^1$ action is invariant under the inversion $u\rightarrow 1/u$, and more importantly, also the topological charge density, therefore
 \be\label{inv-top}
 n(u)=n(1/u).
 \ee
As a consequence, the solution $1/u$ has the same interpretation: it is a system consisting on $n$ lumps, but now located at the poles of $1/u$. The impurity model discussed here does not possess inversion symmetry, that is, in general if $u$ is a solution, $1/u$ is not. This implies, in particular, that the zeroes and poles of the holomorphic and impurity components of the solution do not have a symmetric role. The properties of impurities with $\deg \sigma_1> \deg \sigma_2$ will be studied in Sec. 4.2.

\subsection{Connection with Magnetic Skyrmions}

It was pointed out in the literature \cite{BPS-imp-susy, Schroers, Schroers1} that there is a connection between the $\mathbb{C}P^1$ model with impurities and the Magnetic Skyrmion at critical coupling. The Magnetic Skyrmion is a two dimensional soliton stabilized by a first order term called Dzyaloshinskii-Moriya interaction. This reminds the first order coupling between the field and the impurity which ensures the BPS property. It was shown in \cite{Schroers} that at critical coupling, the BPS Magnetic Skyrmion in terms of the stereographic coordinate $w$ has the following form
\be
w_{MS}=\frac{1}{-\frac{i}{2}\kappa e^{i\alpha}\bar{z}+f(z)},
\ee 
where $\kappa$ and $\alpha$ are real constants of the model and $f(z)$ is a general holomorphic map from the plane to the Riemann sphere. From our previous discussion we have learnt that for a constant impurity of the form
\be
\sigma=-\frac{i}{2}\kappa e^{i\alpha},
\ee
the BPS solutions of the $\mathbb{C}P^1$ impurity model have the form
\be
u_{i\mathbb{C}P^1}=-\frac{i}{2}\kappa e^{i\alpha}\bar{z}+f(z),
\ee
which implies the following relation between the solutions of both models
\be
u_{i\mathbb{C}P^1}=\frac{1}{w_{MS}}.
\ee

In addition, taking into account (\ref{inv-top}), the topological charge density of both solutions coincides and, as a consequence, they describe de same configuration. It is not difficult to envisage an impurity model in the spirit of (\ref{full-lag}) for which its BPS solution coincide exactly with that of the critical Magnetic Skyrmion. The underlying SUSY structure of the model simplifies again the design. First of all, the BPS equation in the off-shell SUSY formulation of the model can be written as
\be
F-2\pa_{\bar{z}} u=0
\ee

The second step is to look for field-impurity couplings such that the algebraic field equation for $F$ gives the correct contribution. For the critical Magnetic Skyrmion this should be
\be
F=-2\sigma u^2.
\ee

This value of the auxiliary field is given by the following terms
\bea
\mathcal{L}_{\mathbb{C}P^1}&=&\frac{1}{2} \frac{ \partial_\mu u \partial^\mu \bar{u}+F\bar{F}}{(1+|u|^2)^2}+\frac{u^2}{(1+\vert u \vert^2)^2}\sigma(x)(F-\pa_{\bar{z}} u)+\text{h.c.}+\text{fermions}.\label{sub:imp2}.
\eea

After solving the $F$ equation, the on-shell Lagrangian has the form
\be
\mathcal{L}=\frac{ 1}{(1+|u|^2)^2}\left(\frac{1}{2}\partial_\mu u \partial^\mu \bar{u} -2\vert u \vert^4\sigma\bar{\sigma}-2u^2\sigma\pa_{\bar{z}}u-2\bar{u}^2\bar{\sigma}\pa_z\bar{u}\right).
\ee

For a constant impurity $\sigma=A\in \mathbb{C}$ the general solution of the BPS equation is given by
\be
\pa_{\bar{z}}u +A u^2=0\Rightarrow u_{i\mathbb{C}P^1}=\frac{1}{A\bar{z}+f(z)},
\ee
which is exactly the BPS solution of the critical Magnetic Skyrmions.

\section{The Moduli space}
\label{moduli}

The manifold defined by the BPS solutions of a given charge is often called the moduli space and denoted by $\mathcal{M}_N$. The low velocity dynamics of solitons belonging to a BPS sector can be approximated by a geodesic motion in  $\mathcal{M}_N$. In the $\mathbb{C}P^1$ case it is well-known that, the coordinates associated to the soliton size cannot be described in $\mathcal{M}_1$ since the associated metric diverges, but it is still possible to restrict the moduli space to describe, for example, only translational degrees of freedom. For $N=2$, scattering of solitons can be described in $\mathcal{M}_2$, even though, due to the conformal invariance, solitons have tendency to collapse or expand indefinitely \cite{Lesse, Ward, Zakrzewski}.  The moduli space of the impurity model shares some similarities with that of the standard $\mathbb{C}P^1$, but it has a reacher structure. In this case, the moduli does not depend only on the topological sector, but also on the choice of the impurity. We will denote the impurity moduli space by  $\mathcal{M}_N^\sigma$. We will not give here a detailed classification of the possible moduli spaces, instead we will analyze the main novelties with respect to the standard case and illustrate them through some examples. 


\subsection{Examples of geodesics}

\begin{figure}
\centering
{
  \includegraphics[width=50mm]{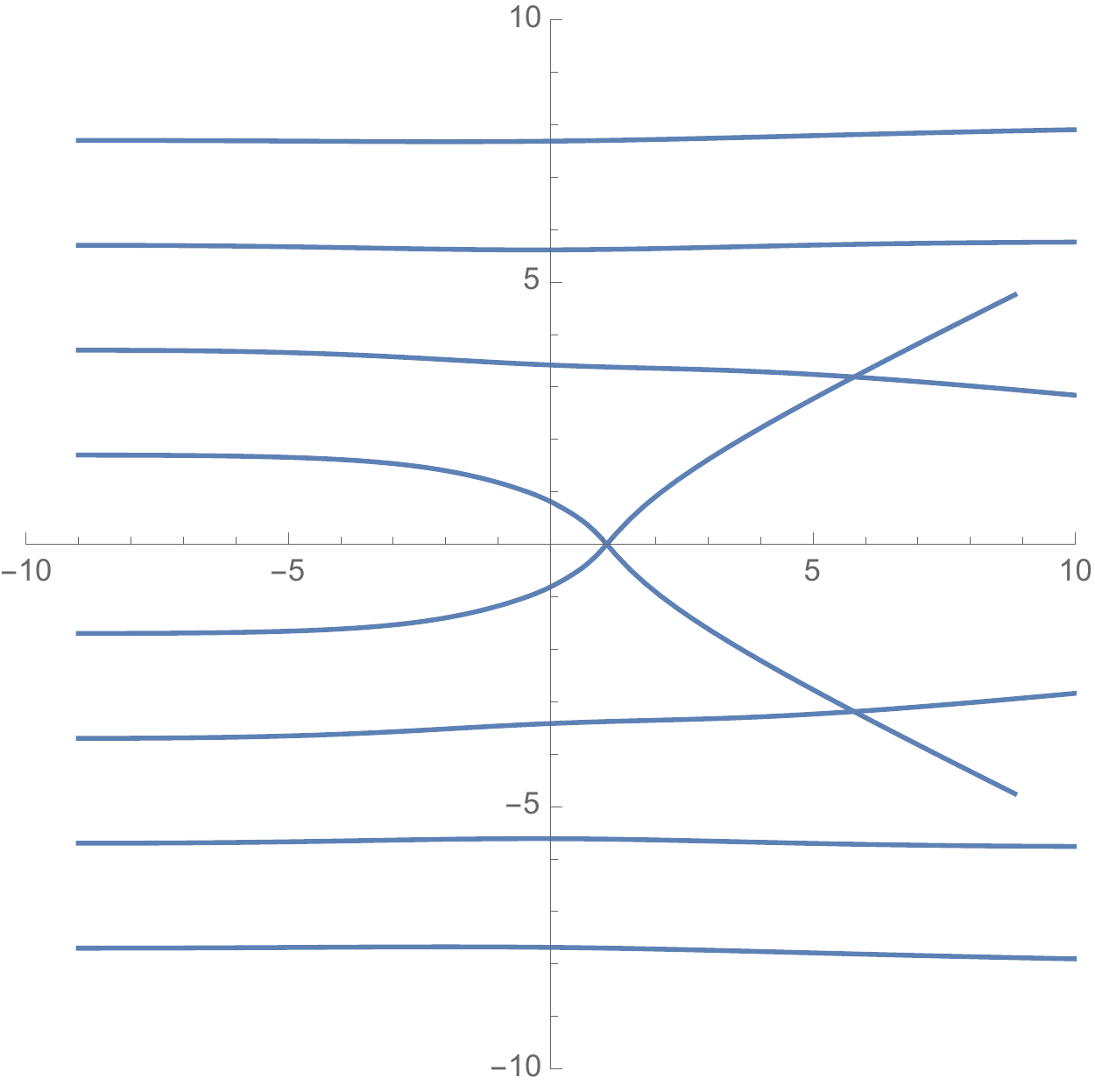}
}
{
  \includegraphics[width=50mm]{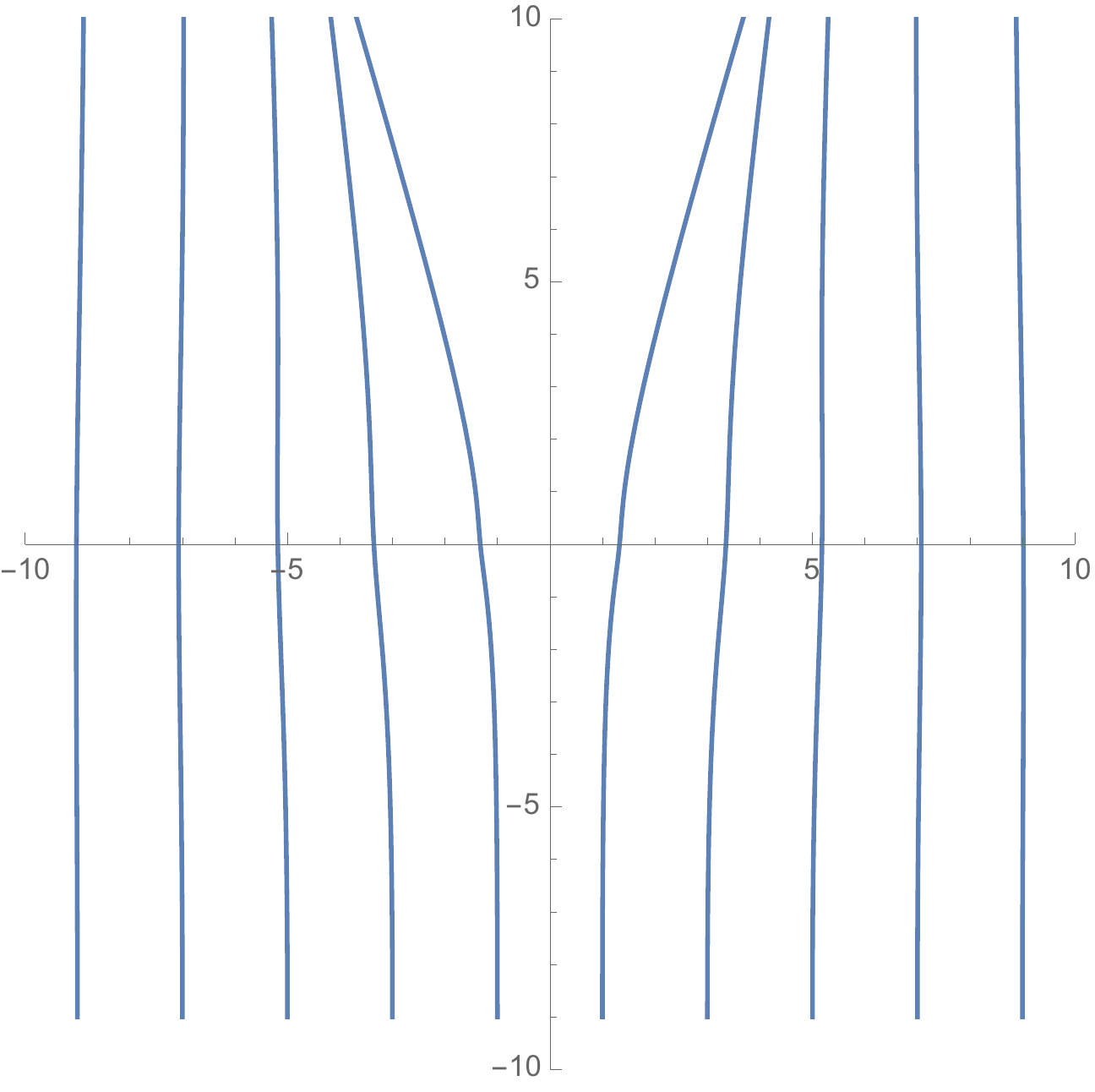}
}
\caption{Geodesics in the regular moduli space of the N=0 charge solutions for $ \beta=2 $. }
\label{moduli_Ncero}
\end{figure}

Let us focus now on the charge $N=0$ configuration with the boundary condition (\ref{boundary}).
As in the usual case without impurity, this base point condition implies a one-to-one correspondence between the charge N lump and the rational holomorphic maps satisfying the condition $\deg p<\deg q$. With this base point condition, the charge 1 solution of the ordinary $\mathbb{C}P^1$ model is given by
\be
u_s=\frac{\beta}{z-\gamma}\label{standard:sol},
\ee
where $\beta\in \mathbb{C}^\star$ and $\gamma\in \mathbb{C}$. It is clear that the moduli space associated to this solution in the standard lump configuration is $\mathcal{M}_1=\mathbb{C}^\star\times \mathbb{C}$. Let us now choose now $ \sigma_2=\bar{z}$, which corresponds to a antilump centered at the origin. The BPS solution in the impurity model corresponding to (\ref{standard:sol}) is
\be
u=\frac{\beta}{z-\gamma}+\frac{1}{\bar{z}}\label{imp:sol}.
\ee
According to the degree formula (\ref{general}) this solution corresponds to an $N=0$ charge in the impurity model and it still satisfies (\ref{boundary}). Moreover, it describes a (dynamical) lump located at $z=\gamma$ and a (non-dynamical) untilump located at the origin. Note that also we have to impose that $\gamma\neq 0$ (so, $q$ and $\sigma_2$ do not have common roots). It is clear that the moduli space of the zero charge solutions of the impurity model $\mathcal{M}_0^\sigma$ verifies $\mathcal{M}_0^\sigma=\mathbb{C}^\star\times \mathbb{C}^\star$. If we allow $\beta$ and $\gamma$ to depend on $t$, the kinetic energy associated with $u$ is given by
\be
T[u]=\int d^2x \frac{1}{2(1+\vert u \vert^2)^2}\vert \dot{u}\vert^2.\label{kinetic}
\ee
As in the standard $\mathbb{C}P^1$ model, the kinetic energy diverges unless $\dot\beta=0$. This effectively reduces the moduli space to $\mathbb{C}^\star$. The configuration (\ref{imp:sol}) corresponds to a lump of (constant) width $\beta$ and centered at $\gamma(t)$ and a antilump fixed at the origin and width $1$. Since (\ref{imp:sol}) is an exact BPS solution, the geodesic motion on the moduli space is a good approximation for small velocities $\dot\gamma\rightarrow 0$ (becoming exact for $\gamma$ constant). The metric on the moduli space can be computed immediately from (\ref{kinetic}). Expanding (\ref{kinetic}) we get
\be
T[u]=\int d^2x \frac{\vert \beta\vert^2 \vert \dot\gamma\vert^2}{\vert z-\gamma\vert^4 (1+\vert u \vert^2)^2}.
\ee
The metric is therefore given by
\be
g_{\gamma\bar{\gamma}}=\int d^2x \frac{\vert \beta\vert^2 }{\vert z-\gamma\vert^4 (1+\vert u \vert^2)^2},
\ee
where $u$ is given by (\ref{imp:sol}). It is worth noting that, even the moduli space of the charge $N=0$ configuration is nontrivial due to the presence of the impurity, see for example Fig. \ref{moduli_Ncero}, for some examples of geodesics in the moduli with $ \beta =2$. The impurity tends to bend the geodesics towards the origin for small impact parameter in the x-direction, while in the y-direction has the opposite effect. As the impact parameter grows, the lump does not feel the impurity and the moduli space becomes trivial. In this situation the geodesics tend to straight lines.

There is another fundamental feature that affects our maps. Let us consider a general solution of the form
\be
u=R(z)+\Sigma(\bar{z}),
\ee
where $R$ and $\Sigma$ are rational maps. Let us assume that $\deg R=\deg \Sigma=n$.  We have
\be
\lim_{z\rightarrow\infty }u= A R^{n}e^{i n \theta}+ B  R^{n}e^{-i n \theta}= A R^n e^{i n \theta}\left(1+\frac{B}{A}e^{-2in\theta}\right) 
\ee
Let us define $B/A=r e^{i t}$. If $n>0$ we may chose the base space condition $u(\infty)=\infty$ almost everywhere, except for $r=1, \, \theta=\frac{t}{2 n}$, where the field vanishes at infinity. As pointed out in \cite{Schroers} in the context of Magnetic Skyrmions, this can be traced back to the fact that such maps do not extend to smooth maps between spheres. Equivalently, this can be also seen from the topological degree jump in the formula (\ref{general}), when the ratio of the leading coefficients of the holomorphic and antiholomorphic parts has modulus one. These solutions, even belonging to the BPS sector, do not have a well-defined moduli. On the other hand, solutions with boundary condition $u(\infty)=0$ (vanishing impurities at infinity) have a well-defined topological degree for all moduli space values.



\subsection{The regularized moduli. Positive degree impurities.}
\label{stable}

One of the limitations of the moduli space approximation for lump configurations comes from the existence of moduli space coordinates with infinite inertia (see \cite{Sutcliffe_bound} for a recent treatment of this problem using boundary metrics). This is clear in the standard situation even for the $N=1$ lump, where a time dependent radius or phase leads to infinite value of the moduli metric for these coordinates. A similar situation arises in the presence of the impurities discussed in Sec. \ref{imp-0-inf}. In order to analyze the origin of this divergence we can expand the BPS solution at infinity
\be
u=\frac{a_1(t)}{z}+\frac{a_2(t)}{z^2}+...+b_1\bar{z}^m+...,
\ee
where we are assuming that the holomorphic part vanishes at infinity and the antiholomorphic contribution with $m\in\mathbb{Z}$ comes from the presence of the impurity. The kinetic energy density associated to $\dot{a}_1(t)$ at infinity is given by
\be
T\vert_{\vert z\vert\rightarrow \infty}\propto
\begin{cases}
\frac{\vert \dot{a}_1(t) \vert^2 }{\vert z\vert^2},\, m\leq 0\\
\frac{\vert \dot{a}_1(t) \vert^2 }{\vert z\vert^{2+4m}},\, m>0
\end{cases}
\ee

Taking into account the integration measure, if $m\leq0$ the kinetic integral gives a logarithmic singularity. However, for $m>0$, the kinetic energy integral associate to the leading term $a_1(t)$ behaves as  $1/r^{4m+2}$, and therefore, it converges.

We conclude that an impurity lump can collapse or expand within the geodesic approximation if the impurity has positive degree. A detailed analysis of this moduli space for low charge solutions is left for a future work.


\subsection{Some solutions and their topological charge densities}
\label{stable}

As we have discussed, the interpretation of the impurity depends crucially on its form. One cannot immediately say that any combination of a holomorphic function (the lump) of degree $n_1$ with a holomorphic function (the impurity) of degree $n_2$ with well-separated zeros (poles) can be seen as a system of $n_1$ lumps and $n_2$ antilumps.

\begin{figure}
\centering
\subfloat[$u_1$, $A=1$, $\gamma=5$ and $\epsilon=-5$]{
  \includegraphics[width=45mm]{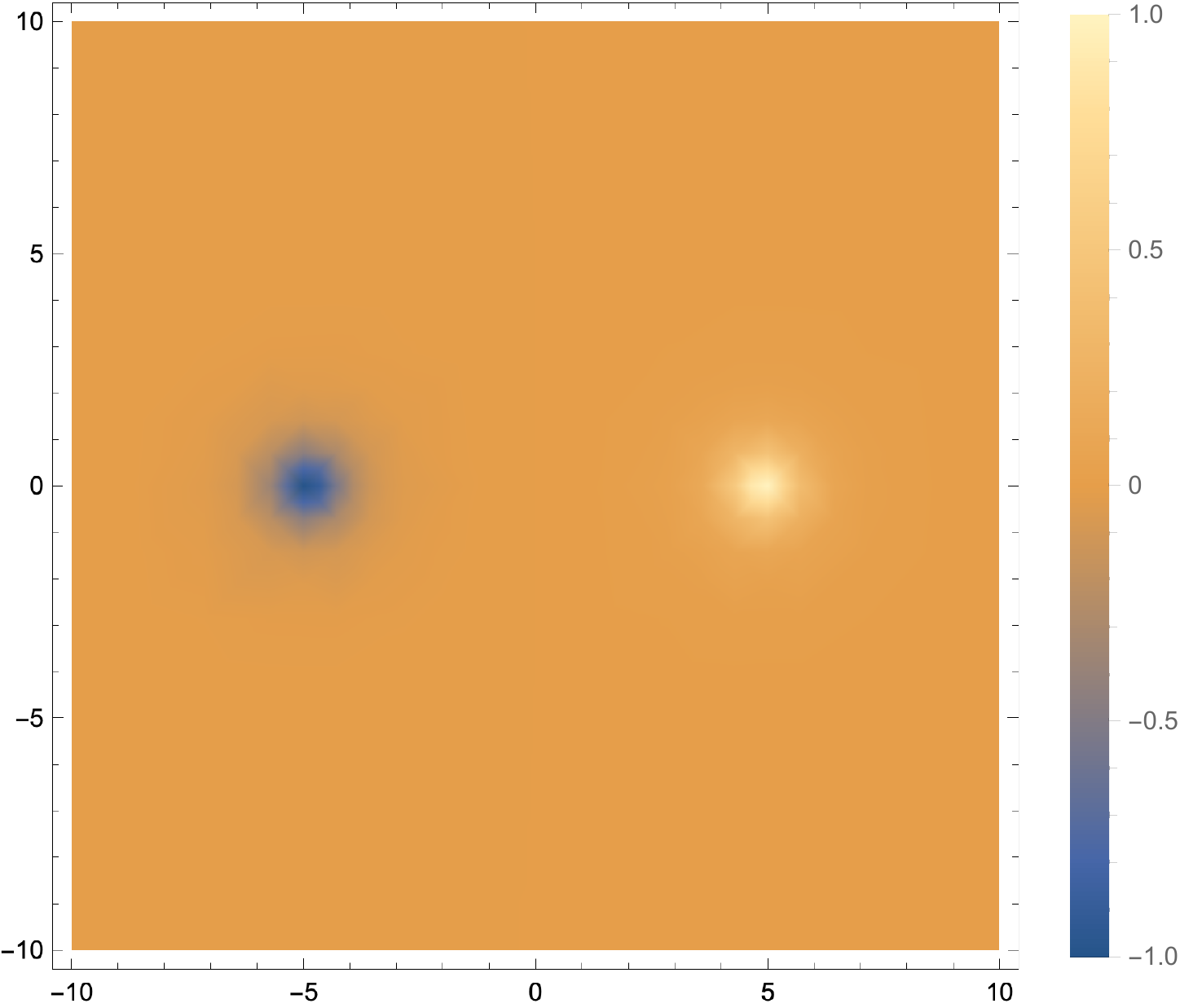}
}
\subfloat[$u_3$, $A=1$, $\gamma=5$ and $\epsilon=-5$]{
  \includegraphics[width=45mm]{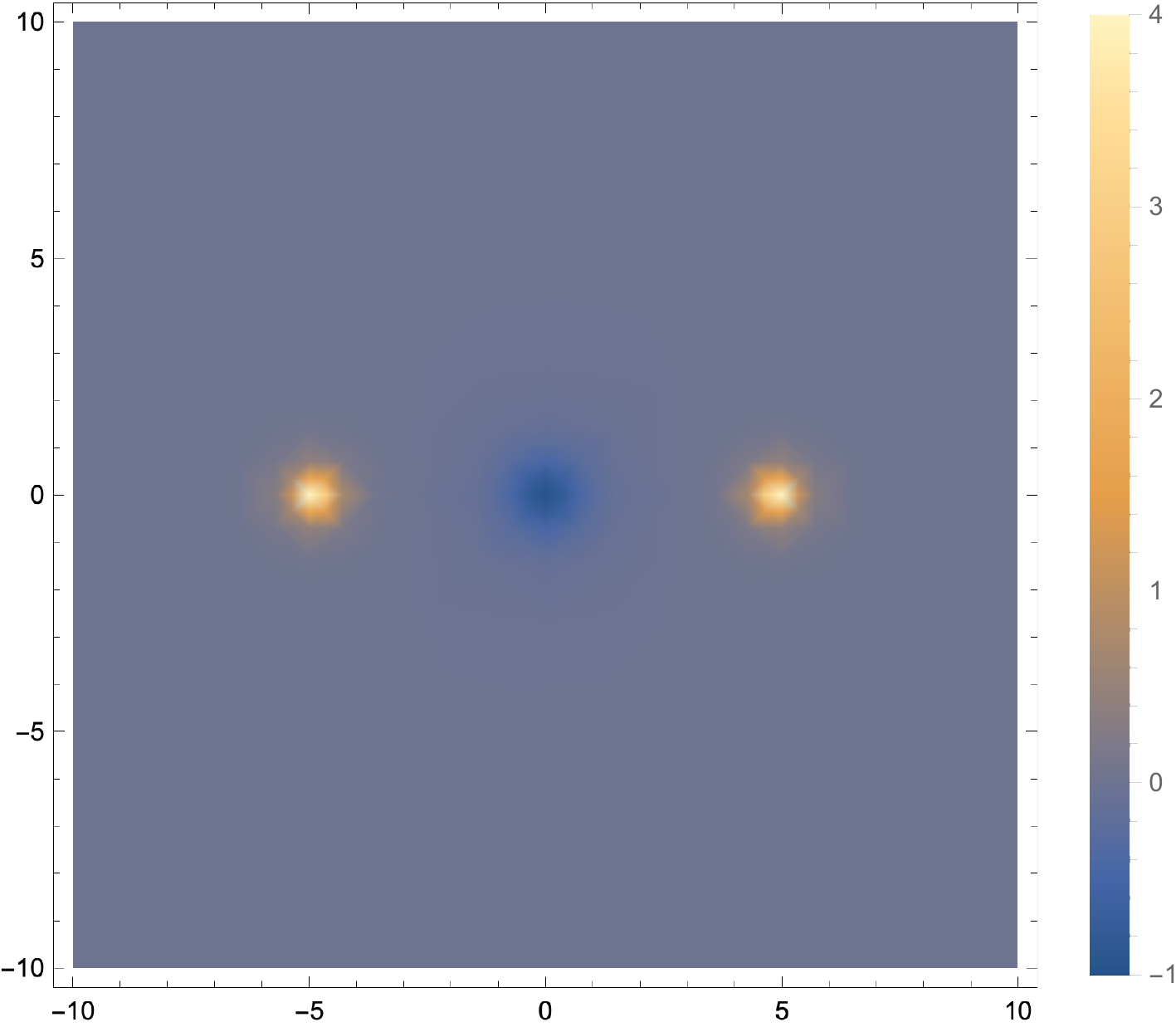}
}
\\
\subfloat[$u_2$, $A=2$ and $\gamma=0$]{
  \includegraphics[width=45mm]{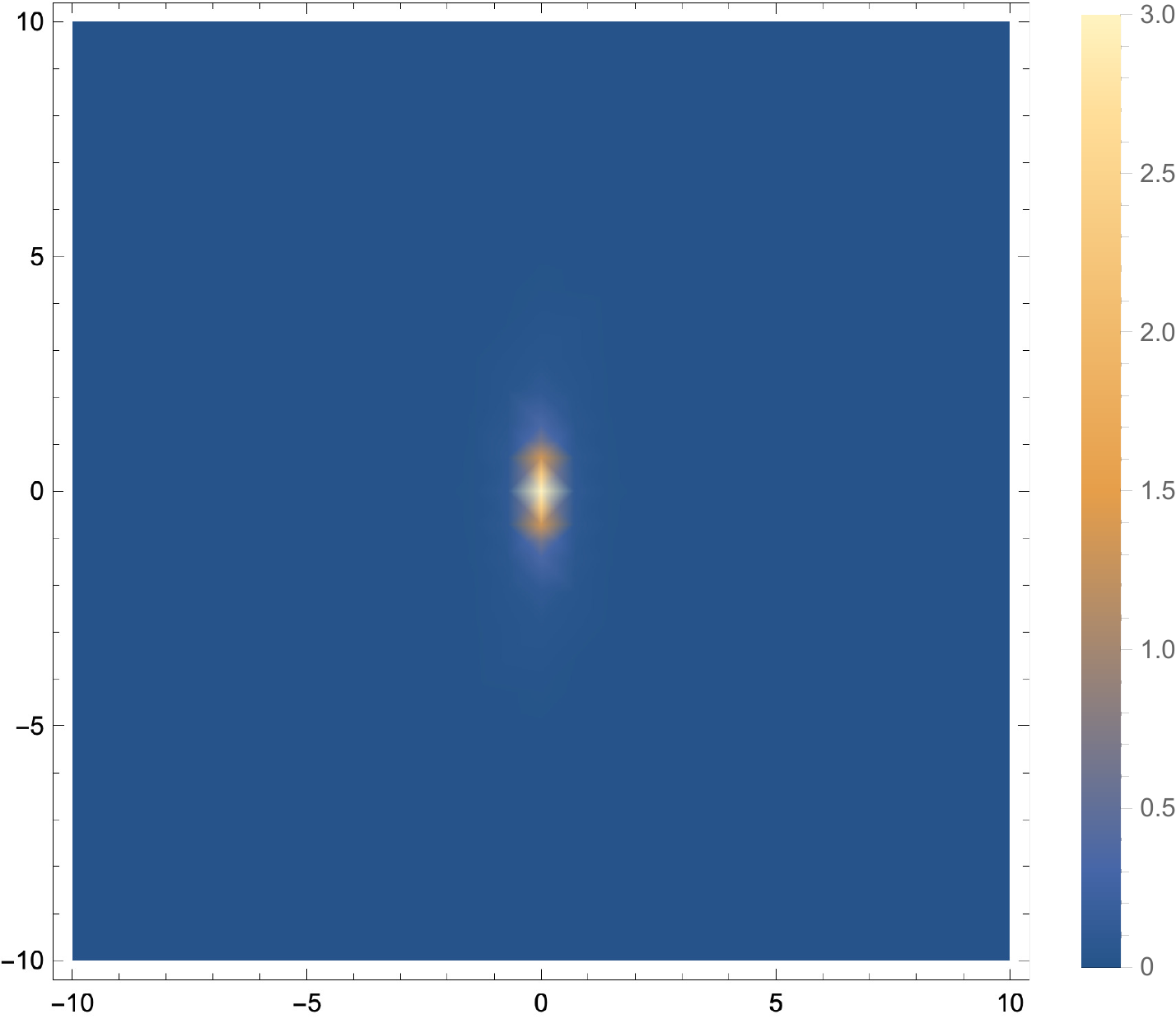}
}
\subfloat[$u_2$, $A=1/2$ and $\gamma=0$]{
  \includegraphics[width=45mm]{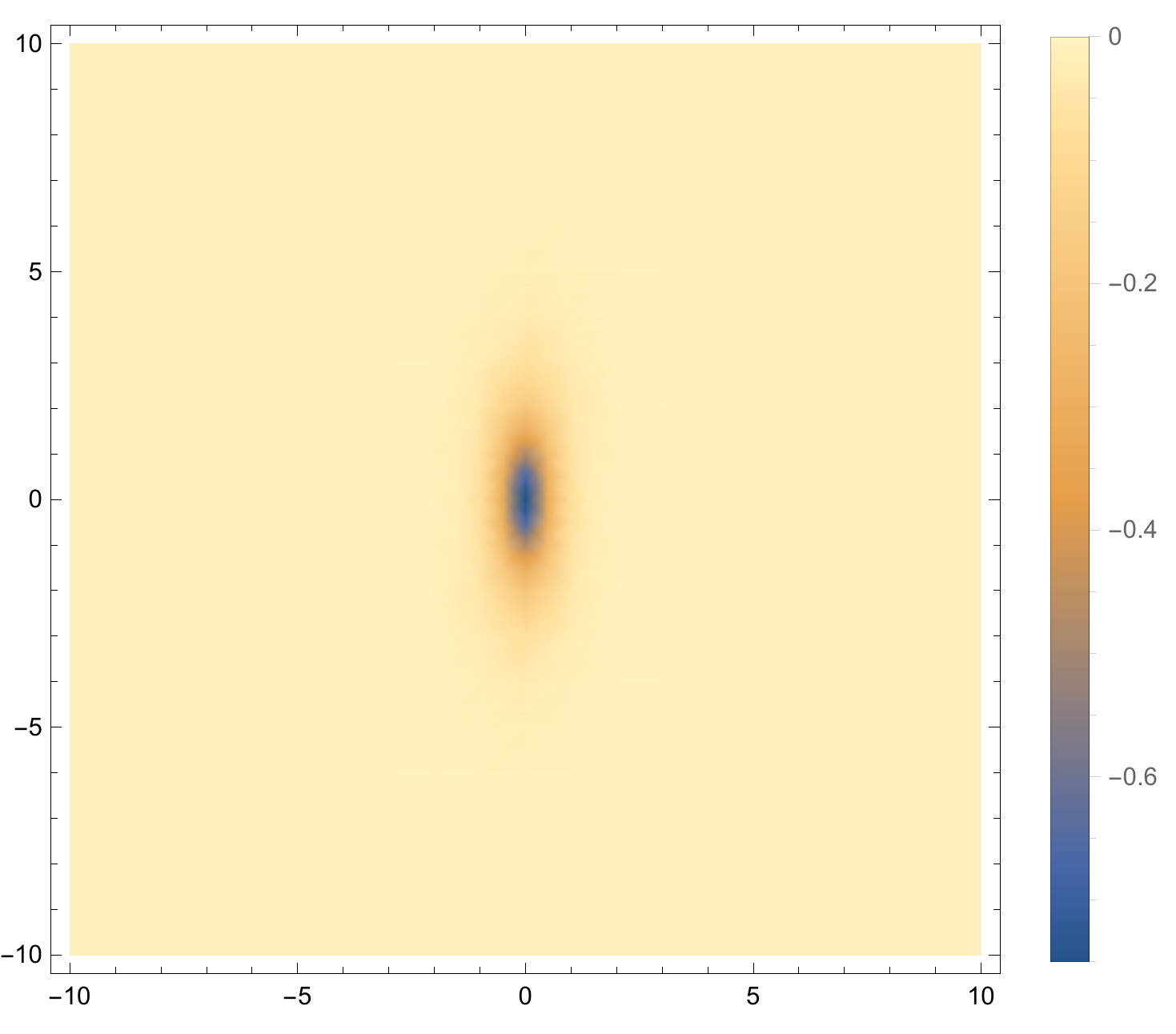}
}
\caption{Behavior of the topological charge density for different solutions. The color palette indicates the topological charge density.}
\label{lumpscatt}
\end{figure}

The first example we consider is the following BPS solution

\be
u_1=\frac{A}{z-\gamma}+\frac{1}{\bar{z}-\epsilon}.
\ee

Our degree formula gives
\be
N(u_1)=0.
\ee

Intuitively, this configuration should correspond to a well-separated lump-antilump system for $\vert\gamma-\epsilon \vert $ large. The lump is centered at $\gamma$, while the antilump is centered at $\epsilon$. This is consistent with the vanishing degree of the configuration and with the considerations of Sec. 3 (see Fig. \ref{lumpscatt} (a)). The second BPS solution we would like to consider is given by
\be\label{sol_change}
u_2=A z-\gamma+\bar{z}
\ee

Despite that it may seem, this solution is more subtle. First, our degree formula gives the following 
\be\label{sol_change_1}
N(u_2)=
\begin{cases}
1,\, \vert A \vert <1\\
0,\, \vert A \vert =1\\
-1,\, \vert A \vert >1
\end{cases}
\ee

When the holomorphic part ``dominates" ($\vert A \vert >1$) the total degree of the solution is $1$, and the full solution can be interpreted as a lump deformed in the $y$-direction (see Fig. \ref{lumpscatt} (c)). In the opposite situation  ($\vert A \vert <1$), the total charge is $-1$, and as a consequence we have a sort of antilump also stretched in the $y$-direction (see Fig. \ref{lumpscatt} (d)). The intermediate situation  ($\vert A \vert =1$), has degree zero. The interpretation of this case is simple, the map $u_2$ with $A=1$ is actually real, and cannot cover the Riemann sphere, it falls therefore in the topological sector of zero charge. 

In order to show the rich variety of solutions we consider finally
\be
u_3=\frac{A z}{(z-\gamma)(z-\epsilon)}+\frac{1}{\bar{z}}
\ee
Once again, the intuitive picture that one can get from the solution is a system of two lumps (at $\gamma$ and $\epsilon$) and one antilump at the origin  (see Fig. \ref{lumpscatt} (b)). This is also consistent with the total charge of the configuration
\be
N(u_3)=1.
\ee

One can say, in general, that when the total degree associated to the impurity (understood as the difference of degrees between numerator and denominator) is negative, the topological degree of the solution is well-defined. That is, if one considers vanishing impurities at infinity, the total degree does not change through the moduli space. Of course, as in the standard lump case, one has to remove also the common root points. On the other hand, if the total impurity degree is positive, the moduli space splits into regions of different degrees.

\section{Kähler property of the regular moduli.}
\label{kahler}

It is the purpose of this section to show an important geometric property of the moduli space of the BPS impurity lump. In the standard situation, the moduli space $\mathcal{M}_N$ of $N$-lump configuration has real dimension $4N$, since it is defined by the set of based rational maps of complex dimension $2N$ .  As we have mentioned, 2 coordinates in the moduli space manifold have infinite inertia. It is known that the restriction to the finite $4N-2$ submanifold, from which we have eliminated the infinite inertia coordinates, is Kähler \cite{Din}. Despite that it may seem, due to the appearance of holomorphic and non-holomorphic coordinates in the BPS solution, we will show here that the moduli space restricted to the finite $4N-2$ submanifold is also Kähler. First, following \cite{Din}, we use the base point condition $u(\infty)=1+B$. The BPS  impurity lump takes the form
\be\label{u_kahler}
u=\frac{p(z)}{q(z)}+B \sigma(\overline{z})=\frac{\prod_{i=1}^N (z-a_i)}{\prod_{i=1}^N (z-b_i)}+B\frac{\prod_{i=1}^m (\overline{z}-\alpha_i)}{\prod_{i=1}^m (\overline{z}-\beta_i)}.
\ee
According to (\ref{general}) this corresponds to a charge $N-m$ solution, and it is well-defined whenever numerators and denominators do not have common roots. We assume now that $a_i\rightarrow a_i(t)$ and  $b_i\rightarrow b_i(t)$ (note that $\alpha_i$ and $\beta_i$ are constant coefficients determined by the choice of the impurity) and compute the kinetic energy
\be
T=\int d^2x \frac{\vert \pa_t u \vert^2}{(1+ \vert u\vert^2)^2}.
\ee

If we introduce
\be
\widetilde{p}=p+B\sigma(\bar{z}) q(z),
\ee
 and taking into account (\ref{u_kahler}) we have
\be
T=\int d^2 x \frac{\vert  \dot{\widetilde{p}}q -\widetilde{p}\dot{q}\vert^2}{(q \overline{q}+\widetilde{p}\overline{\widetilde{p}})^2}=\int d^2 x \frac{\vert  \dot{p}q -p\dot{q}\vert^2}{(q \overline{q}+\widetilde{p}\overline{\widetilde{p}})^2},
\ee
where the term proportional to the impurity has disappeared from the numerator. In order to make explicit the moduli coordinates, the time derivatives can be rewritten as follows
\be\label{kin_2}
T=\int d^2 x \frac{1}{(q \overline{q}+\widetilde{p}\overline{\widetilde{p}})^2}\vert \dot{a}_i \pa_{a_i}-\dot{b}_i\pa_{b_i}\vert^2 \vert p \vert^2 \vert q\vert^2.
\ee

As in the standard case, the idea is to find a Kähler potential from which the metric can be derived. If such potential exist, the Kähler property of the metric is automatically satisfied. After some simple algebraic manipulations we propose the following potential
\be\label{kin-kah}
T=\dot{w}_\mu\dot{\overline{w}}_\nu \int d^2x \pa_{w_\mu} \pa_{\overline{w}_\nu}\log \left( \vert p+B \sigma(\bar{z}) q \vert^2+\vert q\vert^2\right),
\ee
where $u_\mu=a_\mu$ for $\mu=1,..,N$ and $w_\mu=b_{\mu-N}$ for $\mu=N+1,...,2N$. In order to prove the equality between (\ref{kin_2}) and (\ref{kin-kah}) we expand the square in (\ref{kin_2}), split the integral in four terms and compare them with (\ref{kin-kah}). We have
\bea\label{kah_41}
&&\pa_{a_i}\pa_{\bar{a}_i}\log \left( \vert p+B \sigma(\bar{z}) q \vert^2+\vert q\vert^2\right)=\frac{\vert q\vert^2 \pa_{a_i}p\pa_{\bar{a}_i}\bar{p}}{(q \overline{q}+\widetilde{p}\overline{\widetilde{p}})^2},\\\label{kah_42}
&&\pa_{a_i}\pa_{\bar{b}_i}\log \left( \vert p+B \sigma(\bar{z}) q \vert^2+\vert q\vert^2\right)=\frac{\bar{p} q \pa_{a_i}p\pa_{\bar{b}_i}\bar{q}}{(q \overline{q}+\widetilde{p}\overline{\widetilde{p}})^2},\\ \label{kah_43}
&&\pa_{b_i}\pa_{\bar{a}_i}\log \left( \vert p+B \sigma(\bar{z}) q \vert^2+\vert q\vert^2\right)=\frac{p\bar{q}\pa_{b_i}q\pa_{\bar{a}_i}\bar{p}}{(q \overline{q}+\widetilde{p}\overline{\widetilde{p}})^2},\\\label{kah_44}
&&\pa_{b_i}\pa_{\bar{b}_i}\log \left( \vert p+B \sigma(\bar{z}) q \vert^2+\vert q\vert^2\right)=\frac{\vert p\vert^2 \pa_{b_i}q\pa_{\bar{b}_i}\bar{q}}{(q \overline{q}+\widetilde{p}\overline{\widetilde{p}})^2}.
\eea 

By comparing (\ref{kah_41})-(\ref{kah_44}) with (\ref{kin_2}) it is immediate to see the announced equivalence. The expression (\ref{kin-kah}) provides automatically the real Kähler potential 
\be
\mathcal{K}=\int d^2x \log \left( \vert p+B \sigma(\bar{z}) q \vert^2+\vert q\vert^2\right).
\ee

Finally, the Kähler metric has the standard Kähler representation in terms of the potential
\be
g_{\mu\overline{\nu}}=\pa^2_{w_\mu \overline{w}_\nu}\mathcal{K}.
\ee

Several comments are in order. Notice that the only differences w.r.t. the standard case is the addition of extra terms containing the impurity. Also in this case the Kähler potential is strictly divergent, but this divergence does not affect the metric. We have focussed on the $\mathbb{C}P^1$ models, but, at least for two dimensional sigma models with impurities, the Kähler property seems to hold (for a complete proof in general sigma models see \cite{Ruback}).  For the type of impurity we have chosen here (which tends to a nonzero value at infinity) the restriction to the regular moduli is translated to the (standard) constraint $\sum_j (\dot{a}_i-\dot{b}_i)=0$. In principle, if one does not care about the behaviour of the impurity at infinity, while keeping the finite base point condition for the holomorphic part, the form of the Kähler potential is the same (note that we have not used any property of the impurity in our derivation). In addition, if $\lim_{\bar{z}\rightarrow\infty}\sigma(\bar{z})=\infty$, no further constraints are necessary in order to keep the kinetic energy convergent.



\section{Holomorphic impurities as multifield systems.}
\label{gauge}

The interpretation of impurities as frozen lumps pursued throughout this work can be realized by means of a multifield model. 
Let us consider the following Lagrangian
\bea\label{limit_lag}
\mathcal{L}&=&\frac{\Lambda_1}{(1+|u|^2)^2}\left(\frac{1}{2}\partial_\mu u \partial^\mu \bar{u}-2w\bar{w}-2w\pa_{\bar{z}}u-2\bar{w}\pa_z\bar{u}\right)+\frac{\Lambda_2}{(1+|w|^2)^2}\partial_\mu w \partial^\mu \bar{w}.
\eea
This describes two complex fields coupled in the same way as the  BPS impurity model. For $\Lambda_2>>\Lambda_1 $ the fields $u$ and $w$ decouple (for similar results in one-dimensional theories see \cite{multifield}). In this limit, $w$ behaves as a standard lump, therefore
\be\label{limit_imp}
w=\frac{p(z)}{q(z)} \,\, \text{or}\,\, w=\frac{p(\bar{z})}{q(\bar{z})}.
\ee
This gives two branches defined by the holomorphic or antiholomorphic character of $w$. If we choose one antiholomorphic solution and plug it back into the Lagrangian (\ref{limit_lag}), the resulting model in the $\Lambda_1$ sector reproduces the impurity model discussed in the paper, where the impurity now is determine by the $w$-lump. Physically, one may interpret this model as a system of two lump fields such that the energy scale of one of them is much bigger than the other. This effectively freezes one of them, which acts like an impurity which does not back-react to the evolution of the other. 






\section{Conclusions}

We have studied the $\mathbb{C}P^1$ model coupled to impurities in such a way that part of the original self-dual structure is preserved. The Bogomolny bound remains invariant, but the Bogomolny equations are a sort of inhomogenous Cauchy-Rimenann equations augmented by the impurity. Although any impurity coupled to the $\mathbb{C}P^1$ field in the form (\ref{full-lag}) preserves the (half of the) BPS structure, we have restricted our analysis to (anti)holomorphic impurities. This constraint provides a natural interpretation for the impurity: it can be considered as a frozen lump of negative charge, in whose background the positive charge lumps move. This characteristic allows to obtain in a clean way these impurity models as certain limits of theories with two complex fields. 

We have analyzed in detail the topological charge of the BPS solutions and proved that depends crucially on the form of the impurity. Under some circumstances, the topological charge may change the sign through a trajectory in the moduli space, pointing that, certain maps (BPS solutions) do not extend to smooth maps between 2-spheres. This peculiarity was already glimpsed in the context of critical Magnetic Skyrmions \cite{Schroers}. Elaborating on this point, we have clarified the connection between holomorphic BPS impurity models and critical Magnetic Skyrmions, resulting that, in the constant impurity case, they are related through an inversion transformation of the field. This relation implies, in particular, that the topological charge densities of two solutions (related by the inversion) coincide in both models. We have also shown that it is straightforward to build an impurity model whose BPS solution coincide with that of the critical Magnetic Skyrmion case in the limit of constant impurities. 

One feature that moves this family of models away from the standard $\mathbb{C}P^1$ case, is the moduli space structure. It is possible to choose the impurity in such a way that even the moduli coordinate associated to the lump size is well-defined in the geodesic approximation. In the standard case, coordinates with infinite inertia can be related to the conformal invariance of the model. In the impurity model, it does not seem to be the case as conformal symmetry depends on the specific form of the impurity and not only on its behavior at infinity. Remarkably, we have shown that, despite these differences, the geometry of the moduli space is still Kähler. 

Another relevant point is the connection of this family of impurity models with non-impurity models at certain limits. It has been pointed out in \cite{Schroers1} that, at the level of BPS solutions, the $\mathbb{C}P^1$ impurity model can be obtained from a gauge non-abelian ($SU(2)$) $\sigma$-model, where the gauge field is chosen such that the BPS equations of the model mimics those of the BPS impurity. For the rest of the discussion, it is important to emphasize that, in this case, the gauge field itself is taken to be the impurity, i.e., it has no dynamics. A different approach was taken in \cite{tong}. Here, certain types of (magnetic) impurity models were obtained as certain limits of gauge theories with product group $U(1)\times U(1)$. The impurities are interpreted as frozen vortices associated to one of the gauge groups. As a consequence, the impurities are effectively obtained at some limit (in this case the limit of infinite mass of the vortices associated to one of the gauge groups). This is the point of view we take in Sec. 6. We have chosen to couple the would be dynamical lump to  an extra complex field in the same fashion as the field-impurity BPS coupling. In this case, both fields are originally dynamical. Then, the energy scale associated to the second field is taken to be large compared to the other. At this limit, both fields decouple and the BPS equation of the ``heavier" one forces the holomorphic character of the solution. Once this solution is reintroduced in the action, acts as an impurity,  in whose background the other moves. This provides a natural interpretation of impurities as frozen (anti)lumps. 

As we have mentioned the impurity models discussed here possess an underlying SUSY structure. This structure can be used to build new impurity models even in the higher-derivative case \cite{queiruga_susy_1, queiruga_susy_2, queiruga_susy_3, nita_1, nita_2, nita_3}. Further research, regarding the study of the fermionic sectors of the SUSY impurity models as well as generalizations to other spacetime dimensions, are left for a future work.



{\bf Acknowledgements.}- This work is supported by the Spanish Ministry MCIU/AEI/FEDER grant (PGC2018-094626-B-C21) and the Basque Government grant (IT-979-16). The author is grateful to A. Wereszczynski for useful comments. 

 
\renewcommand{\theequation}{A.\arabic{equation}}
\setcounter{equation}{0}

\section*{Appendix A: Subtleties in the definition of the impurity topological degree} \label{A1}

The topological degree of a map $u$ between Riemann spheres can be written as follows
\be
N=\frac{i}{2 \pi}\int \frac{du\wedge d\bar{u}}{\left(1+\vert u\vert^2\right)^2},\label{degree1}
\ee
where $u$ is the map given by
\be
u=\frac{p(z)}{l(z)}+\frac{\sigma_1(\bar{z})}{\sigma_2(\bar{z})}=\frac{p(z)\sigma_2(\bar{z})+ \sigma_1(\bar{z}) l(z)}{l(z)\sigma_2(\bar{z})}\equiv \frac{r(z,\bar{z})}{q(z,\bar{z})}
\ee
We assume that the polynomials $p(z),\,l(z)$ and $\sigma_1(\bar{z}),\sigma_2(\bar{z})$ do not have common roots and their degrees are $n, k, s_1$ and $s_2$ respectively. It can be shown easily that
\be
\frac{du\wedge d\bar{u}}{\left(1+\vert u\vert^2\right)^2}=\frac{1}{2}\,d\left(\frac{u d\bar{u}-\bar{u} du}{1+\vert u\vert^2} \right).\label{dec1}
\ee

We follow the procedure of \cite{Schroers}.  The expression in the brackets of the r.h.s of (\ref{dec1}) can be expanded as follows 
\be
\frac{u d\bar{u}-\bar{u} du}{1+\vert u\vert^2}=\frac{r d\bar{r}-\bar{r} dr+q d\bar{q}-\bar{q}dq}{r\bar{r}+q\bar{q}}+d\ln q-d\ln\bar{q}.\label{expan1}
\ee
Since $p$ and $r$ do not have common roots, it follows that the first term in the r.h.s. of (\ref{expan1}) is non-singular. The second term, however, has poles at the zeros of $q(z,\bar{z})$. We can use now the Stokes' theorem for forms. If $\omega$ is a one-form and $\Omega$ is a region in the complex plane we have
\be
\int_{\pa\Omega} \omega=\int_\Omega d\omega.\label{stokes}
\ee

Let $\Gamma_\eta^i$ be a circle of radius $\eta$ surrounding the zero $\beta_i$ of $q(z,\bar{z})$ and $\Gamma_R$ a circle of radius $R$ centered at the origin. Taking into account (\ref{dec1}), (\ref{expan1}) and (\ref{stokes}), the integral (\ref{degree1}) over the complex plane can be written as
\be
N=\frac{i}{4\pi}\lim_{\eta\rightarrow 0}\sum_{i=1}^k\int_{\Gamma_\eta^i}\left(d\ln q-d\ln\bar{q}\right)+\frac{i}{4\pi}\lim_{R\rightarrow\infty}\int_{\Gamma_R}\frac{u d\bar{u}-\bar{u} du}{1+\vert u\vert^2}\equiv I_1+I_2  \label{degree2}
\ee
The first term in the r.h.s. of (\ref{degree2}) ($I_1$) can be computed using the residue theorem. It produces a positive contribution from the holomorphic zeroes of $l(z)$ and a negative contribution from the antiholomorphic zeroes of $\sigma_2$, that is $I_1=k-s_2$. For the second term ($I_2$) we define $z=R e^{  i \theta}$, expand for $R\rightarrow \infty$ and integrate. We have three cases:\\

Case i): $n-k>s_1-s_2$. The integral $I_2$ gives

\be
I_2=
\begin{cases}
n-k,\, n>k \\
0,\, n\leq k 
\end{cases}
\ee

Case ii): $n-k<s_1-s_2$. The integral $I_2$ gives

\be
I_2=
\begin{cases}
s_2-s_1,\, s_1>s_2 \\
0,\, s_1\leq s_2 
\end{cases}
\ee

Case iii): $n-k=s_1-s_2$. Assume that the leading coefficient of the rational function $p(z)/l(z)$ at $z$ is $A\in \mathbb{C}$ ,and the leading coefficient of $s_1(\bar{z})/s_2(\bar{z})$ is $B\in \mathbb{C}$.  Let us assume that $s_1>s_2,$, we have three possibilities depending on the modulus of $B/A$. If we define $B/A=r e^{i t}$ and $m=s_1-s_2$ we have
\be
I_2(r,t)=-\frac{m(r^2-1)}{2 \pi}\int_0^{2\pi}\frac{d\theta}{1+r^2-2 r \cos(t-2 m \theta)}
\ee

For $m>0$ and $r=1$ we have $I_2(r,t)=0$. For $m>0$ and $r>1$, $I_2(r,t)=-m$. Now, since $I_1(r,t)=-I_2(1/r,t)$ we have for $m>0$ and $r<1$, $I_2(r,t)=m$. For $m\leq 0$, $I_2=0$. Therefore, finally
\be
N=
\begin{cases}
n-s_2,\, n-k>s_1-s_2, \, n>k \\
k-s_2,\, n-k>s_1-s_2, \, n\leq k  \\
k-s_1,\, n-k<s_1-s_2,\, s_1> s_2\\
k-s_2,\, n-k<s_1-s_2, \, s_1 \leq s_2 \\
k-s_1,\, n-k=s_1-s_2,\, s_1>s_2,\, r>1\\
n-s_2,\,n-k=s_1-s_2,\, s_1>s_2,\, r<1\\
k-s_2,\,n-k=s_1-s_2,\, s_1>s_2,\, r=1\\
k-s_2,\,n-k=s_1-s_2,\, s_1\leq s_2
\end{cases}\label{general}
\ee

Note that the degree is well-defined, except in the case where the degree of the leading monomial of both rational functions coincide at infinity with positive degree. In this case, the degree of the solutions depends on the ratio of the leading coefficients.

\end{document}